\documentclass{ws-mpla}
\usepackage{textcomp}
\usepackage[super]{cite}
\def\Journal#1#2#3#4{{#1} {\bf #2}, #3 (#4)}

\def\NIMA{{\em Nucl. Instrum. Methods} A}

\def\PLB{{\em Phys. Lett.}  B}
\def\PRL{\em Phys. Rev. Lett.}
\def\PRD{{\em Phys. Rev.} D}

\def\GaC{\em Gravitation and Cosmology}

\def\JETPL{\em JETP Lett.}

\def\CQG{\em Class. Quantum Grav.}
\def\APJ{\em Astrophys. J.}

\def\IJMPA{{\em Int. J. Mod. Phys.}  A}
\def\IJMPD{{\em Int. J. Mod. Phys.}  D}

\def\AIPCP{\em AIP Conf. Proc.}

\def\JPCS{{\em J. Phys.:} Conf. Ser.}
\def\BWP{\em Bled Workshops in Physics}

\def\keV{\,{\rm keV}}
\def\MeV{\,{\rm MeV}}

\def\({\left(}
\def\){\right)}

\def\beq{\begin{equation}}
\def\eeq{\end{equation}}
\def\bea{\begin{eqnarray}}
\def\eea{\end{eqnarray}}

\begin{document}

\markboth{J.R. CUDELL, M.Yu. KHLOPOV, Q. WALLEMACQ}
{EFFECTS OF DARK ATOM EXCITATION}

\catchline{}{}{}{}{}

\title{EFFECTS OF DARK ATOM EXCITATIONS}
\author{JEAN-RENE CUDELL}

\address{IFPA, D\'ep. AGO, Universit\'e de Li\`ege, Sart Tilman, 4000 Li\`ege, Belgium\\
jr.cudell@ulg.ac.be}

\author{MAXIM YU. KHLOPOV}

\address{National Research Nuclear University "MEPHI" (Moscow Engineering Physics Institute) and \\
    Centre for Cosmoparticle Physics "Cosmion" 115409 Moscow, Russia \\
APC laboratory 10, rue Alice Domon et L\'eonie Duquet \\75205
Paris Cedex 13, France\\
khlopov@apc.univ-paris7.fr}

\author{QUENTIN WALLEMACQ}

\address{IFPA, D\'ep. AGO, Universit\'e de Li\`ege, Sart Tilman, 4000 Li\`ege, Belgium\\
quentin.wallemacq@ulg.ac.be}

\maketitle

\pub{Received (Day Month Year)}{Revised (Day Month Year)}

\begin{abstract}
New stable quarks and charged leptons may exist and be hidden from detection,
as they are bound by Coulomb interaction in neutral dark atoms of composite dark
matter.
This possibility leads to fundamentally new types of indirect effects related to
the excitation of such dark atoms followed by their electromagnetic de-excitation.
Stable -2 charged particles O$^{--}$, bound to primordial helium in O-helium
(OHe) atoms, represent the simplest model of dark atoms. Here we consider
the structure of OHe atomic levels which is a necessary input for the indirect
tests of such composite dark matter scenarios, and we give the spectrum of
electromagnetic transitions from the levels excited in OHe collisions.

\keywords{Elementary particles; composite dark matter; dark atoms; energy levels;
cosmic electromagnetic radiation;
indirect effects of dark matter.}
\end{abstract}

\ccode{PACS Nos.: 95.35.+d, 97.80.Jp, 95.85.Pw}

\section{Introduction}
According to modern cosmology, dark matter corresponds to
$\sim 25\%$ of the total cosmological density, is nonbaryonic and
consists of new stable particles. Such particles (see e.g. references \refcite{DMRev} and \refcite{DADM}
for a review) should
be stable, saturate the measured dark matter density and decouple
from plasma and radiation at least before the beginning of the matter-dominated
era. The easiest way to satisfy these conditions is to
involve neutral elementary weakly interacting massive particles (WIMPs). However
it is not the only particle physics solution for the dark matter
problem and more elaborate models of composite dark matter are
possible. The simplest model of dark atoms
is the O-helium (OHe) model, in which stable -2 charge particles O$^{--}$ are
bound by the Coulomb interaction with primordial helium\cite{DMRev,DADM}.

Elementary particle frameworks for heavy stable -2 charged species are
provided by: (a) stable ``antibaryons'' $\bar U \bar U \bar U$ formed
by anti-$U$ quark of fourth
generation\cite{Q,I,lom,KPS06,Belotsky:2008se,Khlopov:2006dk};
(b) AC-leptons\cite{Khlopov:2006dk,5,FKS,Khlopov:2006uv}, predicted in an
extension\cite{5} of the standard model, based on the approach of
almost-commutative geometry\cite{bookAC};  (c) Technileptons and
anti-technibaryons\cite{KK} in the framework of walking technicolor
models
(WTC)\cite{Sannino:2004qp,Hong:2004td,Dietrich:2005jn,Dietrich:2005wk,Gudnason:2006ug,Gudnason:2006yj}.

If new stable species belong to non-trivial representations of
the electroweak SU(2) group, sphaleron transitions at high temperatures
can relate the baryon asymmetry to the excess of
-2 charge stable species, as it was demonstrated in the case of WTC
in references \citenum{KK,Levels1,KK2,unesco,iwara,I2}. The only free parameter in this case is
the mass of O$^{--}$.

 After it is formed
in the Standard Big Bang Nucleosynthesis (SBBN), $^4$He screens the excessive
O$^{--}$ charged particles in composite ($^4$He$^{++}$O$^{--}$) O-helium (OHe)
``atoms''\cite{I}.
In all the considered forms of O-helium, O$^{--}$ behaves either as a lepton or
as a heavy quark cluster with strongly suppressed hadronic
interaction. Therefore the O-helium interaction with matter is
determined by the nuclear interactions of He. This neutral primordial
nuclear-interacting species can play the role of a nontrivial form of strongly
interacting dark
matter\cite{Starkman,Wolfram,Starkman2,Javorsek,Mitra,Mack,McGuire:2001qj,McGuire2,ZF},
giving rise to a Warmer-than-Cold dark-matter scenario\cite{Levels,Levels1,KK2}.
It should be noted that the
nuclear cross section of the O-helium
interaction with matter escapes the severe
constraints\cite{McGuire:2001qj,McGuire2,ZF}
on strongly interacting dark matter particles
(SIMPs)\cite{Starkman,Wolfram,Starkman2,Javorsek,Mitra,Mack,McGuire:2001qj,McGuire2,ZF}
imposed by the XQC experiment\cite{XQC,XQC1}.

Slowed down in the terrestrial matter OHe collisions cause negligible nuclear
recoil in the underground detectors, thus avoiding severe constraints of direct
dark matter searches in CDMS\cite{cdms}, XENON100\cite{xenon} and LUX\cite{lux}
experiments. It makes indirect effects of OHe especially important to test
OHe hypothesis.

Here we concentrate on the structure of OHe atoms, their excitation in collisions and the observable effects of electromagnetic radiation from their de-excitation.

\section{The atomic structure of OHe}\label{spectrum}

\begin{figure}
\begin{center}
\includegraphics[scale=0.7]{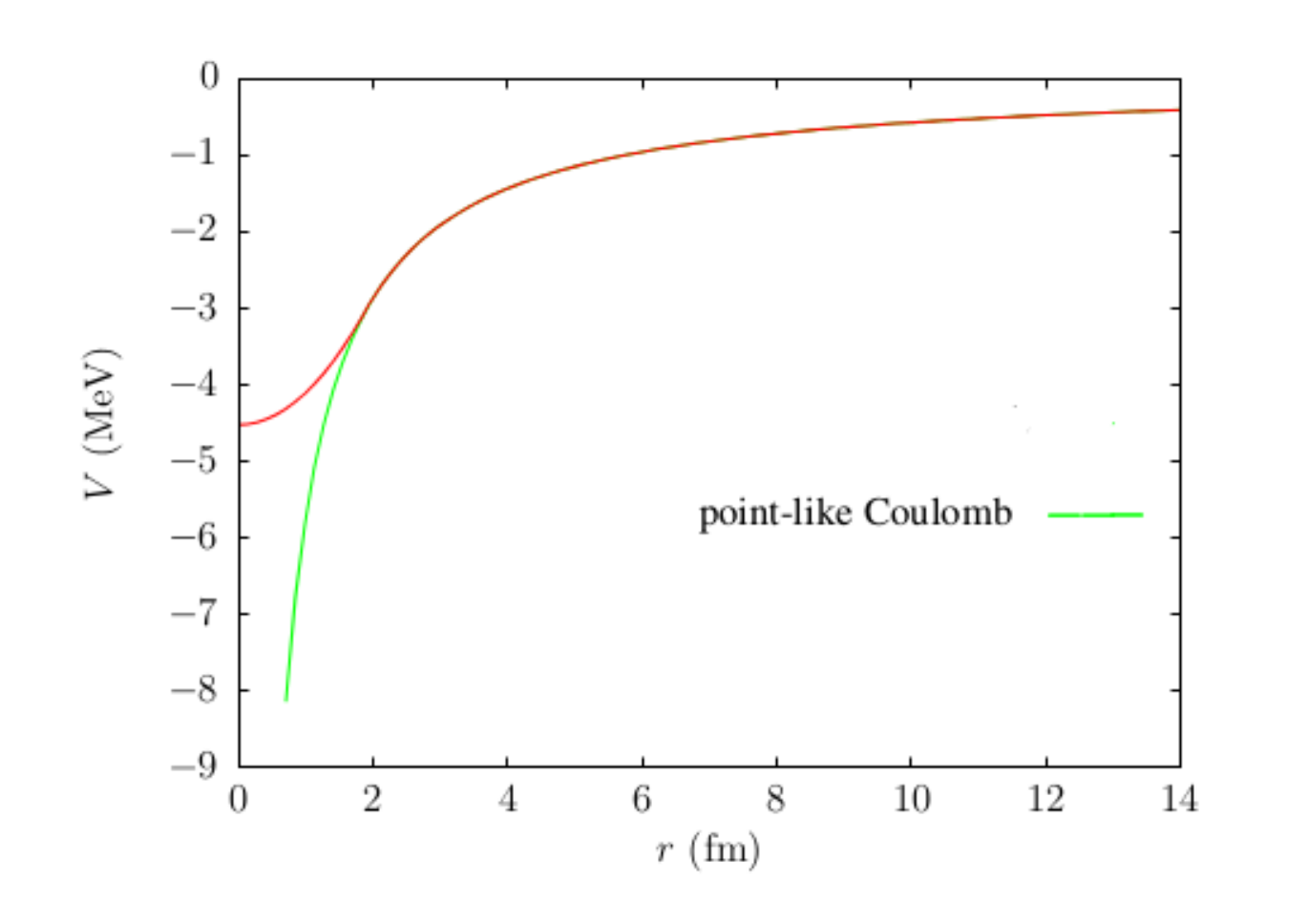}
\end{center}
\caption{The O-helium interaction potential (red) and the elementary Coulomb potential (green) obtained for a point-like helium nucleus, as a function of the distance $r$ between O$^{--}$ and the center of the helium nucleus.}
\label{pot}
\end{figure}

An OHe atom is made of a heavy O$^{--}$ particle and a helium nucleus, of
respective masses $M_O$ and $M_{He}$ such that $M_O\gg M_{He}$, bound by the
Coulomb interaction in a hydrogen-like structure. While the O$^{--}$ particle, of
charge $Z_O=2e$, is point-like, we approximate the charge distribution
of the helium nucleus as a uniformly
charged sphere of radius $R_{He}=1.2\times 4^{1/3}$~fm. This allows us to take
 the O-helium
attractive interaction potential as that between a point and a sphere, i.e.:
\begin{equation}
\begin{aligned}
V(r\geq R_{He}) & = -\frac{Z_O Z_{He}\alpha}{r} \\
V(r<R_{He}) & =-\frac{Z_O Z_{He}\alpha}{2R_{He}}\left(3-\frac{r^2}{R_{He}^2}\right),
\end{aligned}
\label{potential}
\end{equation}
where $\alpha=e^2/4\pi$ is the fine structure constant and $r$ is the distance
between O$^{--}$ and the center of the helium nucleus. The potential is
represented in Figure \ref{pot} together with the elementary Coulomb potential
obtained
when the helium nucleus is assumed to be point-like.

To obtain the eigenstates of an OHe atom, we solve the radial time-independent
Schr\"odinger equation with the potential \eqref{potential}, written in the
center-of-mass frame of the system, which corresponds almost exactly to the
O$^{--}$ particle as $M_O\gg M_{He}$. At zero angular momentum $l$, we use the
non-modified WKB approximation as the potential $V$ is regular at the origin
($\lim\limits_{r \to 0}rV(r)=0$). For the states at higher $l$, we use the
modified WKB approximation, which consists in replacing $l(l+1)$ by $(l+\frac{1}
{2})^2$ in the centrifugal term of the effective potential\cite{froman}.

The eigenvalues $E_{n,l}$ are shown in Table \ref{Table1} for the first ten values
of the principal quantum number $n$ and the corresponding angular momenta
$l=0,...,n-1$, together with the pure hydrogen-like solutions $E^H_n=-\frac{1}
{2}M_{He}\frac{(Z_O Z_{He}\alpha)^2}{n^2}$. We notice a lift of degeneracy on $l$
because the potential \eqref{potential} is no longer $\propto 1/r$, so that the
energy levels now depend on both $n$ and $l$. Also, the energy of the ground state
$1s$ has been increased significantly with respect to $E^H_1$, due to the
finiteness of the well at $r=0$ when the charge distribution of the helium nucleus
is taken into account. In general, at fixed $n$, the pure hydrogen-like energy
levels constitue a lower limit to which the levels $E_{n,l}$ tend as $l$
increases, i.e. as the states are excited and thus as O$^{--}$ and He lie further
apart from each other, making the helium nucleus increasingly point-like. At very
high $n$ the differences between both types of levels are very small and we
recover the pure hydrogen-like case when $n\rightarrow \infty$. As it is well
known that the WKB approximation is less accurate for the deeper bound states, we
computed the energy of the ground state by a variational method using up to $11$
hydrogen-like $s$-orbitals and found the result $-1.1771$ MeV, which shows that
the error on the values given in Table \ref{Table1} is less than 1
\textperthousand.
\begin{table}
\tbl{Energy levels $E_{n,l}$ (MeV) of the OHe atom, for the first ten values of the principal quantum number $n$ and the corresponding angular momenta $l=0,...,n-1$. In the last column are also shown the pure hydrogen-like solutions $E^H_n$ (MeV) obtained when the helium nucleus is assumed to be point-like. The exponents indicate the power of $10$ by which the numbers have to be multiplied to obtain the energy in MeV.}{
\begin{tabular}{c|c|c|c|c|c|c}
\hline
\hline
 $n$ & $l=0$ & $l=1$ & $l=2$ & $l=3$ & $l=4$ & $E^H_n$ \\
\hline
\hline
1 & -1.1760  & - & - & - & - & -1.5879 \\
2 & -0.3446  & -0.3969 & - & - & - & -0.3970 \\
3 & -0.1607  & -0.1764 & -0.1764 & - & - & -0.1764 \\
4 & -9.2538$^{-2}$  & -9.9230$^{-2}$ & -9.9240$^{-2}$ & -9.9239$^{-2}$ & - & -9.9244$^{-2}$ \\
5 & -6.0057$^{-2}$  & -6.3511$^{-2}$ & -6.3511$^{-2}$ & -6.3511$^{-2}$ & -6.3510$^{-2}$ & -6.3516$^{-2}$\\
6 & -4.2097$^{-2}$ & -4.4106$^{-2}$ & -4.4106$^{-2}$ & -4.4106$^{-2}$ & -4.4106$^{-2}$  & -4.4108$^{-2}$  \\
7 & -3.1136$^{-2}$ & -3.2404$^{-2}$ & -3.2404$^{-2}$ & -3.2404$^{-2}$ & -3.2404$^{-2}$ &  -3.2406$^{-2}$\\
8 & -2.3957$^{-2}$ & -2.4808$^{-2}$ & -2.4811$^{-2}$ & -2.4810$^{-2}$ & -2.4810$^{-2}$ & -2.4811$^{-2}$ \\
9 & -1.9002$^{-2}$ & -1.9602$^{-2}$ & -1.9602$^{-2}$ & -1.9602$^{-2}$ & -1.9602$^{-2}$ & -1.9604$^{-2}$ \\
10 & -1.5439$^{-2}$ & -1.5878$^{-2}$ & -1.5878$^{-2}$ & -1.5878$^{-2}$ & -1.5878$^{-2}$ & -1.5879$^{-2}$\\
 \hline
 \hline
  $n$ & $l=5$ & $l=6$ & $l=7$ & $l=8$ & $l=9$ & $E^H_n$ \\
\hline
\hline
 1 & -  & - & - & - & - & -1.5879 \\
2 & - & - & - & - & - & -0.3970 \\
3 & -  & - & - & - & - & -0.1764 \\
4 & -  & - & - & - & - & -9.9244$^{-2}$ \\
5 & -  & - & - & - & - & -6.3516$^{-2}$\\
6 & -4.4105$^{-2}$ & - & - & - & - & -4.4108$^{-2}$  \\
7 & -3.2403$^{-2}$ & -3.2406$^{-2}$ & - & - & - &  -3.2406$^{-2}$\\
8 & -2.4810$^{-2}$ & -2.4810$^{-2}$ & -2.4810$^{-2}$ & - & - & -2.4811$^{-2}$ \\
9 & -1.9604$^{-2}$ & -1.9602$^{-2}$ & -1.9602$^{-2}$ & -1.9602$^{-2}$ & - & -1.9604$^{-2}$ \\
10 & -1.5878$^{-2}$ & -1.5878$^{-2}$ & -1.5878$^{-2}$ & -1.5878$^{-2}$ & -1.5879$^{-2}$ & -1.5879$^{-2}$ \\
  \hline
 \hline
\end{tabular}
\label{Table1}}
\end{table}
\section{Some signatures of O-helium excitation}
\subsection{Positron annihilation line in the galactic bulge}
\begin{figure}
\begin{center}
\includegraphics[scale=0.7]{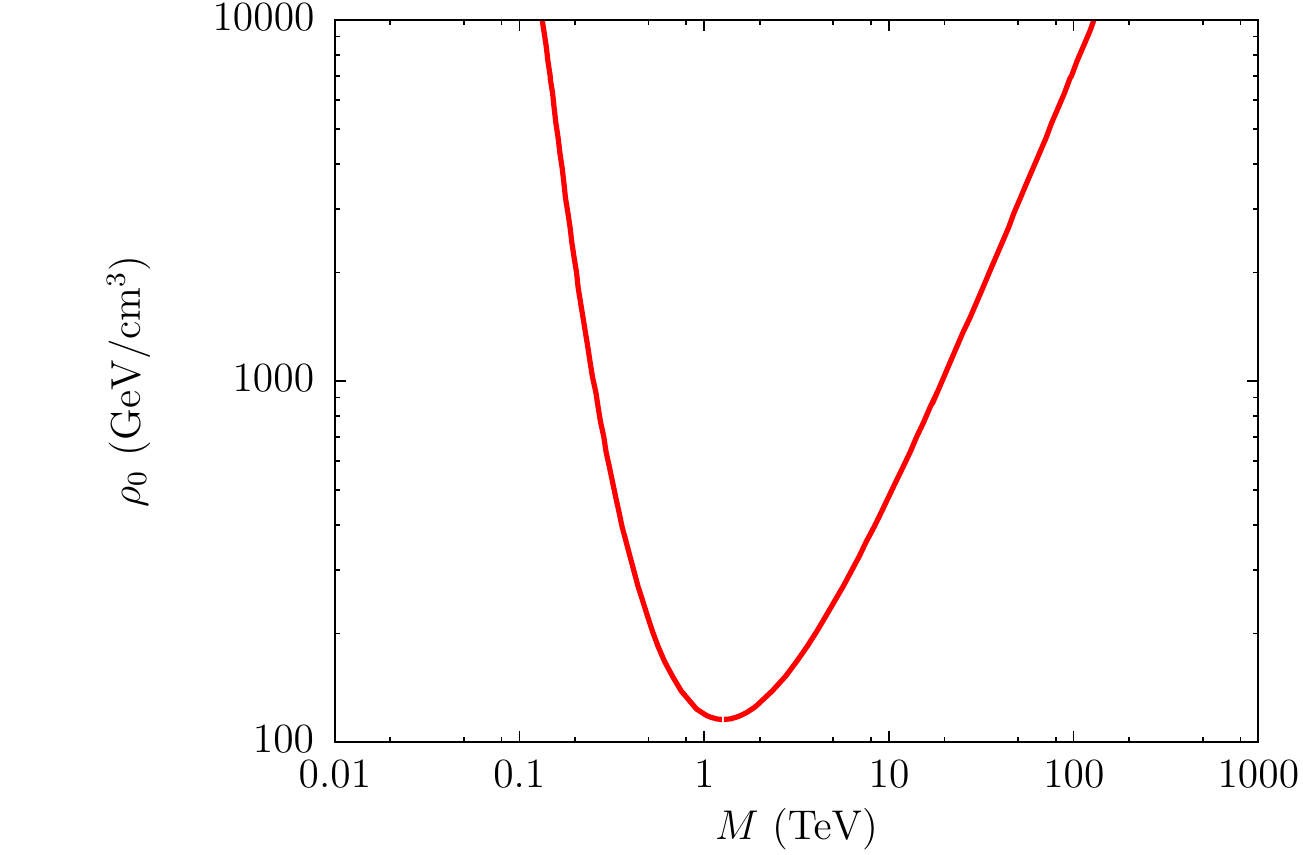}
\end{center}
\caption{Values of the central dark matter density $\rho_{0}$ (GeV/cm$^3$), assuming a Burkert profile, and of the OHe mass $M_O$ (TeV) reproducing the excess of $e^{+} e^{-}$ pair production from the galactic bulge. Below the red curve, the predicted rate is too low.}
\label{rho0_MOHe_1S2S}
\end{figure}

It was first noted in ref.~\citenum{I2} that OHe collisions in the galactic bulge
can lead to OHe de-excitation by pair production, which can provide
a positron-production rate sufficient\cite{Finkbeiner:2007kk} to explain the
excess in the positron annihilation line from the bulge measured by INTEGRAL (see
ref.~\refcite{integral} for a review and references). Indeed, if the 2S level is
excited in an OHe collision, pair production dominates over the two-photon channel
as the de-excitation can go through an E0 transition, so that positron production
is not accompanied by a strong gamma signal. A more detailed analysis of this
possibility\cite{CKW} has shown that the rate of positron production strongly
depends on the velocity and density profiles of OHe in the center of the Galaxy
and a range of parameters was found for which this rate is sufficient to explain
the INTEGRAL data. These results are shown in Figure \ref{rho0_MOHe_1S2S} taken
from ref. \refcite{CKW}.
\subsection{O-helium de-excitation}
It was noted in ref. \refcite{I2} that if OHe levels with non-zero angular
momentum are excited though collisions, gamma lines should appear from
E1 transitions from levels with principal quantum numbers $n$ and $m$ with
$ n>m$, which would have energies
$E_{nm}= 1.5879 \MeV (1/m^2-1/n^2)$ for hydrogen-like OHe, or from the similar
transitions corresponding
to the more realistic case discussed in Section \ref{spectrum}.
 These predictions may be of interest for the analysis of the possible nature of
 unidentified lines, observed in the center of the Galaxy. The lines with energies
 above 20 keV can be searched for in the INTEGRAL data, while a forest of lines of
 lower energy may be found in X-ray observations. In particular, high-level
 transitions can lead to a number of lines around 3.5 keV, which can be
 checked in the XMM-Newton data which are sensitive to the range $0.1-12$ keV.

\begin{figure}
\begin{center}
\includegraphics[scale=0.7]{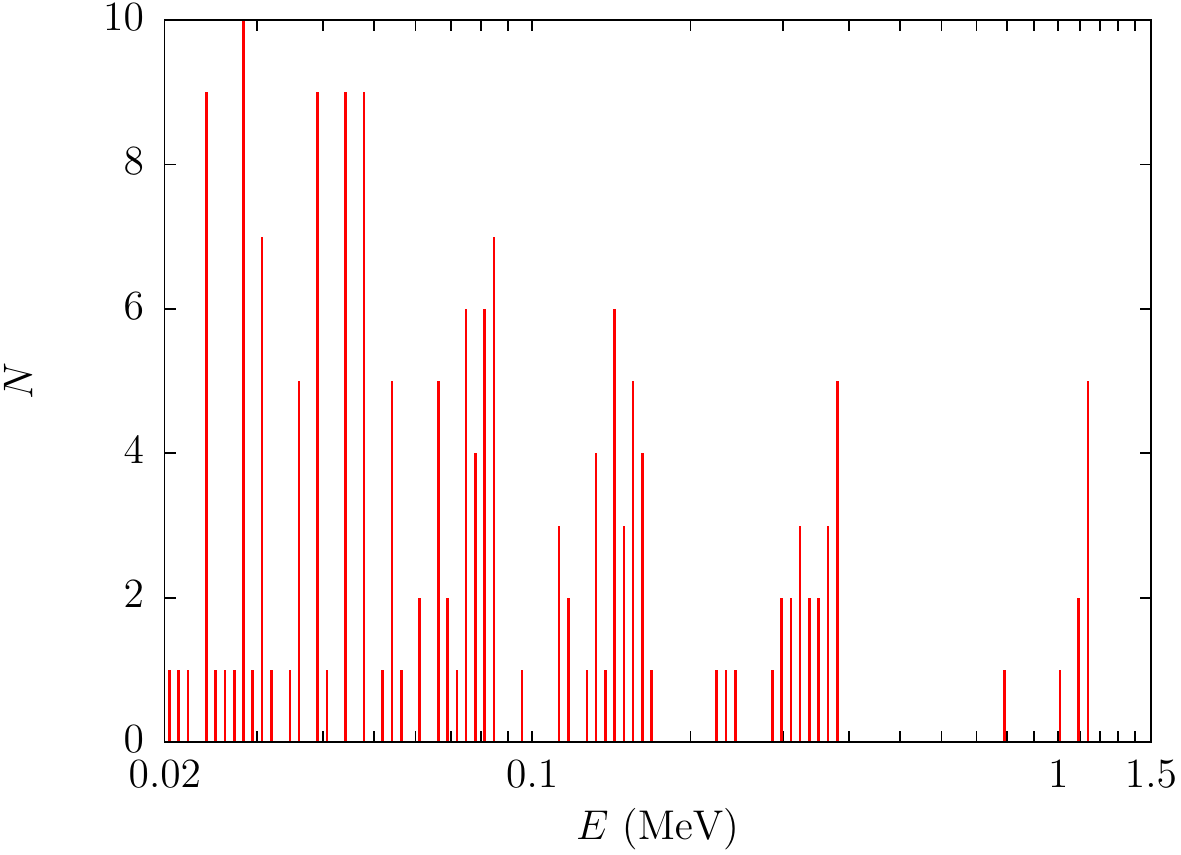}
\end{center}
\caption{The number $N$ of E1 transitions with energies contained within logarithmic
bins in energy of width $\Delta \log(E/20 \keV)=\log(1162 \keV/ 20 \keV)/100$,
as a function of the energy $E$ of the transition.}
\label{density}
\end{figure}

By considering all the possible electric dipole transitions (E1) between the
states presented in Table \ref{Table1}, we find several hundreds of allowed lines,
with energies from the eV range to the MeV range, even for the limited sample
$n\leq 10$. In
Figure \ref{density}, we show the number $N$ of E1 transitions as a function of
the energy $E$, from $20$ keV to $1.162$~MeV, the latter being the largest energy
of the sample and corresponding to the transition from the state $(n=10,l=1)$ to
the ground state. This was obtained by counting the number of E1 transitions with
energies contained within logarithmic bins in energy of width $\Delta
\log (E)=\log (1162 \keV/20 \keV)/100$. The lines with energies between
$3$ and $4$ keV are listed in Table \ref{Table2}.
The comparison of our predictions with the observations provides an effective tool
to test the OHe composite-dark-matter model.

\begin{table}
\tbl{E1 transitions of an OHe atom with energies between $3$ and $4$ keV.}{
\begin{tabular}{c|c|c}
\hline
\hline
Initial state $(n,l)$ & Final state $(n',l')$ & Energy (keV) \\
\hline
\hline
$(5,0)$ & $(5,1)$ & 3.4542 \\
\hline
$(10,1)$ & $(9,0)$ & 3.1236 \\
         & $(9,2)$ & 3.7243 \\
\hline
$(10,2)$ & $(9,1)$ & 3.7243 \\
         & $(9,3)$ & 3.7241 \\
\hline
$(10,3)$ & $(9,2)$ & 3.7245 \\
         & $(9,4)$ & 3.7242 \\
\hline
$(10,4)$ & $(9,3)$ & 3.7243 \\
         & $(9,5)$ & 3.7258 \\
\hline
$(10,5)$ & $(9,4)$ & 3.7246 \\
         & $(9,6)$ & 3.7243 \\
\hline
$(10,6)$ & $(9,5)$ & 3.7260 \\
         & $(9,7)$ & 3.7245 \\
\hline
$(10,7)$ & $(9,6)$ & 3.7240 \\
         & $(9,8)$ & 3.7240 \\
\hline
$(10,8)$ & $(9,7)$ & 3.7245 \\
\hline
$(10,9)$ & $(9,8)$ & 3.7233 \\
\hline
\hline
\end{tabular}}
\label{Table2}
\end{table}

\section{Conclusions}
The existence of heavy stable particles is one of the popular solutions for the
dark matter problem.
These particles are usually considered to be electrically neutral, but dark matter
can also
be formed by
stable heavy charged particles bound in neutral atom-like states by the ordinary
Coulomb attraction.
The analysis of the cosmological evolution and atomic composition of the Universe
constrains the particle charge to be $-2$. These doubly charged
constituents will be trapped by primordial helium
in neutral O-helium states, and this can avoid the problem of overproduction of
 anomalous isotopes of chemical elements, which are severely constrained by
 observations. O-helium dark matter
might explain puzzles of direct dark matter searches\cite{DDMRev}. However, such
an explanation implies specific properties of the OHe-nucleus interaction, which
 may not be realized\cite{Bled}. Then OHe can
be elusive for direct dark matter searches, making its indirect effect of special
interest.

The present work sheds new light on the indirect effects of
dark matter. Specific electromagnetic signatures of dark atoms are challenging for
the experimental searches and are, possibly, already observed as the excess of the
positron annihilation line radiation from the galactic bulge in the INTEGRAL data.
Confrontation of the predicted spectrum of OHe de-excitation with the cosmic X-ray
and gamma ray data provides a new way for an indirect probe of the composite dark
matter hypothesis.

\section*{Acknowledgments}
The research of J.R.C. and Q.W. was supported by the Fonds de la Recherche Scientifique - FNRS under grant 4.4501.05. and Q.W. is supported by the Fonds de la Recherche Scientifique - FNRS as a Research Fellow. The part of work of M.Kh. related with various forms of dark matter was supported by the grant RFBR 14-22-03048 and the other part of his work by the Ministry of Education and Science of Russian Federation, project 3.472.2014/K

\end{document}